# ACHIEVEMENTS AND LESSONS FROM TEVATRON*

V. Shiltsev[#], FNAL, Batavia, IL 60510, U.S.A.

*Abstract*

For almost a quarter of a century, the Tevatron proton-antiproton collider was the centerpiece of the world's high energy physics program – beginning operation in December of 1985 until it was overtaken by LHC in 2011. The aim of the this unique scientific instrument was to explore the elementary particle physics reactions with center of mass collision energies of up to 1.96 TeV. The initial design luminosity of the Tevatron was $10^{30}$cm$^{-2}$s$^{-1}$, however as a result of two decades of upgrades, the accelerator has been able to deliver 430 times higher luminosities to each of two high luminosity experiments, CDF and D0. Tevatron will be shut off September 30, 2011. The collider was arguably one of the most complex research instruments ever to reach the operation stage and is widely recognized for many technological breakthroughs and numerous physics discoveries. Below we briefly present the history of the Tevatron, major advances in accelerator physics, and technology implemented during the long quest for better and better performance. We also discuss some lessons learned from our experience.

## HISTORY AND PERFORMANCE

The Tevatron was conceived by Bob Wilson [1] to double the energy of the Fermilab complex from 500 GeV to 1000 GeV. The original name, the "Energy Saver/Doubler", reflected this mission and the accrued benefit of reduced power utilization through the use of superconducting magnets. The introduction of superconducting magnets in a large scale application allowed the (now named) Tevatron to be constructed with the same circumference of 6.3 km, and to be installed in the same tunnel as the original Main Ring proton synchrotron which would serve as its injector (at 150 GeV). Superconducting magnet development was initiated in the early 1970's and ultimately produced successful magnets, leading to commissioning of the Tevatron in July 1983.

In 1976 D.Cline et al. proposed a proton-antiproton collider with luminosities of about $10^{29}$ cm$^{-2}$sec$^{-1}$ at Fermilab [2] or at CERN, based on the conversion of an existing accelerator into a storage ring and construction of a new facility for the accumulation and cooling of approximately $10^{11}$ antiprotons per day. The motivation was to discover the intermediate vector bosons. The first antiproton accumulation facility was constructed at CERN and supported collisions at 630 GeV(center-of-mass) in the modified SPS synchrotron, where the W and Z particles were discovered in 1983.

Meanwhile, in 1978 Fermilab decided that proton-antiproton collisions would be supported in the Tevatron, at a center-of-mass energy of 1800 GeV and that an Antiproton Source facility would be constructed to supply the flux of antiprotons needed for design luminosity of $1\times10^{30}$ cm$^{-2}$sec$^{-1}$.

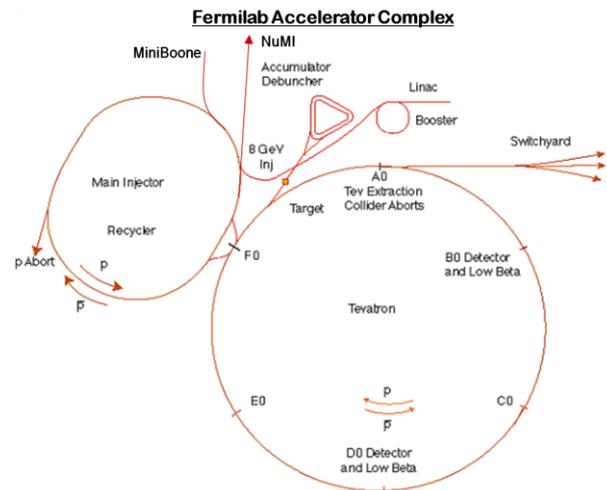

Figure 1: Layout of the Fermilab accelerator complex

The Tevatron as a fixed target accelerator was completed in 1983 [3]. The Antiproton Source [4] was completed in 1985 and first collisions were observed in the Tevatron using operational elements of the CDF detector (then under construction) in October 1985. Initial operations of the collider for data taking took place during a period from February through May of 1987. A more extensive run took place between June 1988 and June 1989, representing the first sustained operation at the design luminosity. In this period of operation a total of 5 pb$^{-1}$ were delivered to CDF at 1800 GeV (center-of-mass) and the first western hemisphere W's and Z's were observed. The initial operational goal of $1\times10^{30}$ cm$^{-2}$sec$^{-1}$ luminosity was exceeded during this run. Table I summarizes the actual performance achieved in the 1988-89 run. (A short run at √s = 1020 GeV also occurred in 1989.)

In the early to mid-1990's a number of improvements were implemented to prepare for operation of Collider Run I (August of 1992 through February 1996):

<u>Electrostatic Separators</u> aimed at mitigating the beam-beam limitations by placing protons and antiprotons on separate helical orbits, thus allowing an increase in the number of bunches and proton intensity: twenty-two, 3 m long, electrostatic separators operating at up to ±300 kV across a 5 cm gap were installed into the Tevatron by 1992. During Run II (2001-2011), 4 additional separators were installed to improve separation at the nearest parasitic crossings.

___

Low beta systems which ultimately allowed operations with $\beta^*$ less than 30: The 1988-89 Run did not have a matched insertion for the interaction region at B0 (where CDF was situated). Two sets of high performance quadrupoles were developed and installed at B0 and D0 (which came online for Run I in 1992).

Cryogenic cold compressors to lower the operating Helium temperature by about 0.5 K, thereby allowing the beam energy to be increased to 1000 GeV, in theory. In operational practice 980 GeV was achieved.

Antiproton Source Improvements: A number of improvements were made to the stochastic cooling systems in the Antiproton Source in order to accommodate higher antiproton flux generated by continuously increasing numbers of protons on the antiproton production target. Improvements included the introduction of transverse stochastic cooling into the Debuncher and upgrades to the bandwidth of the core cooling system. These improvements supported an accumulation rate of $7 \times 10^{10}$ antiprotons per hour in concert with the above listed improvements.

Run I consisted of two distinct phases, Run Ia which ended in May of 1993, and Run Ib which was initiated in December of 1993. The 400 MeV linac upgrade (from the initial 200 MeV) was implemented between Run Ia and Run Ib with the goal of reducing space-charge effects at injection energy in the Booster and provide higher beam brightness at 8 GeV. At the result, the total intensity delivered from the Booster increased from roughly $3 \times 10^{12}$ per pulse to about $5 \times 10^{12}$. This resulted in more protons being transmitted to the antiproton production target and, ultimately, more protons in collision in the Tevatron.

Run I ultimately delivered a total integrated luminosity of 180 pb$^{-1}$ to both CDF and D0 experiments at $\sqrt{s}$ = 1800 GeV. By the end of the run the typical luminosity at the beginning of a store was about $1.6 \times 10^{31}$ cm$^{-2}$sec$^{-1}$, a 60% increase over the Run I goal. (A brief colliding run at $\sqrt{s}$ = 630 GeV also occurred in Run I.)

In preparation for the next Collider run, construction of the Main Injector synchrotron and Recycler storage ring was initiated and completed in the spring of 1999 with the Main Injector initially utilized in the last Tevatron fixed target run.

The Main Injector was designed to significantly improve antiproton performance by replacing the Main Ring with a larger aperture, faster cycling machine [5]. The goal was a factor of three increase in the antiproton accumulation rate (to $2 \times 10^{11}$ per hour), accompanied by the ability to obtain 80% transmission from the Antiproton Source to the Tevatron from antiproton intensities up to $2 \times 10^{12}$. An antiproton accumulation rate of $2.5 \times 10^{11}$ per hour was achieved in Collider Run II, and transmission efficiencies beyond 80% for high antiproton intensities were routine.

The Recycler was added to the Main Injector Project midway through the project (utilizing funds generated from an anticipated cost under run.) As conceived, the Recycler would provide storage for very large numbers of antiprotons (up to $6 \times 10^{12}$) and would increase the effective production rate by recapturing unused antiprotons at the end of collider stores [6]. The Recycler was designed with stochastic cooling systems but R&D in electron cooling was initiated in anticipation of providing improved performance. Antiproton intensities above $5 \times 10^{12}$ were ultimately achieved although routine operation was eventually optimized around $4 \times 10^{12}$ antiprotons. Recycling of antiprotons was never implemented.

Table 1: Achieved performance parameters for Collider Runs I and II (typical values at the beginning of a store.)

|  | 1988-89 Run | Run Ib | Run II |  |
|---|---|---|---|---|
| Energy (c.o.m.) | 1800 | 1800 | 1960 | GeV |
| Protons/bunch | 7.0 | 23 | 29 | $\times 10^{10}$ |
| Antiprotons/bunch | 2.9 | 5.5 | 8.1 | $\times 10^{10}$ |
| Bunches/beam | 6 | 6 | 36 |  |
| Total Antiprotons | 17 | 33 | 290 | $\times 10^{10}$ |
| P-emittance (rms, n) | 4.2 | 3.8 | 3.0 | $\pi$ μm |
| Pbar emittance(rms,n) | 3 | 2.1 | 1.5 | $\pi$ μm |
| $\beta^*$ | 55 | 35 | 28 | cm |
| Luminosity (typical) | 1.6 | 16 | 350 | $10^{30}$ cm$^{-2}$s$^{-1}$ |
| Luminosity Integral | 5·10$^{-3}$ | 0.18 | 11.9 | fb$^{-1}$ |

The Main Injector (MI) and Recycler (RR) completed the Fermilab accelerator complex development - see the ultimate scheme of operational accelerators in Fig.1 - and constituted the improvements associated with Collider Run II [7]. The luminosity goal of Run II was $8 \times 10^{31}$ cm$^{-2}$sec$^{-1}$, a factor of five beyond Run I. However, incorporation of the RR into the Main Injector Project was projected to provide up to an additional factor of 2.5.

Run II was initiated in March of 2001 and continued through September 2011. A number of difficulties were experienced in the initial years of operations. These were ultimately overcome through experience accumulated in the course of operation and the organization and execution of a "Run II Upgrade Plan". At the end of the Run II, typical Tevatron luminosities were well in excess of $3.4 \times 10^{32}$ cm$^{-2}$ sec$^{-1}$, with record stores exceeding $4.3 \times 10^{32}$ cm$^{-2}$sec$^{-1}$ – see achieved performance parameters in Table I.

The Collider performance history (see Fig.2) shows the luminosity increases occurred after numerous improvements, some were implemented during operation and others were introduced during regular shutdown periods. They took place in all accelerators and addressed all parameters affecting luminosity – proton and antiproton intensities, emittances, optics functions, bunch length, losses, reliability and availability, etc. Analysis [8] indicates that as the result of some 32 major improvements in 2001-2011, the peak luminosity has grown by a factor of about 54 from $L_i \approx 8 \times 10^{30}$ cm$^{-2}$s$^{-1}$ to

$L_f \approx 430 \times 10^{30}$ cm$^{-2}$s$^{-1}$, or about 13% per step on average. The pace of the Tevatron luminosity progress was one of the fastest among high energy colliders [9]. Further details of the accelerator complex evolution, basic operation of each machine and luminosity performance can be found in [8] and references therein. A detailed account of the first decades of the Fermilab's history can be found in the book [10].

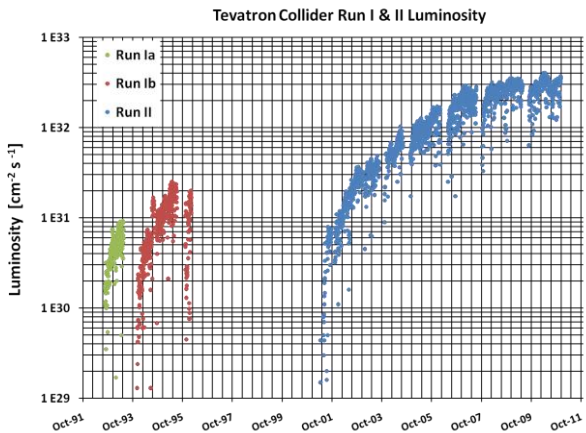

Figure 2: Initial luminosity for all Collider stores

## MAJOR ACCELERATOR PHYSICS AND TECHNOLOGY ACHIEVMENTS

Below we give just few examples of numerous achievements in the field of accelerator technology and beam physics which were incepted and implemented during more than three decades of the Tevatron history. For more details, readers can be referred to the list of references at the end of this article, to articles published in the *JINST Special Issue* [11], or to the book [12].

*Tevatron Superconducting Magnets:*
Superconducting magnets define the Tevatron, the first synchrotron built with the technology [13]. The Tevatron SC magnets experience paved the way for other colliders: HERA, RHIC, LHC. Issues that had to be addressed included conductor strand and cable fabrication, coil geometry and fabrication, mechanical constraint and support of the coils, cooling and insulation, and protection during quenches. The coil placement, and hence magnetic field uniformity at the relative level of few $10^{-4}$, had the biggest effect on the accelerator performance. The magnets which were designed in the 70's (Fig.3) performed beautifully over the years, though offered us a number of puzzles to resolve for optimal operation, like "chromaticity snap-back" effect [14] and coupling due to the cold-mass sagging [15].

*Recycler Permanent Magnets:*
Recycler was the first high energy accelerator ever built with permanent magnets. It also was arguably the cheapest accelerator built (per GeV) and it employs 362 gradient dipole and 109 quadrupole magnets made of SrFe (peak field of about 1.4T) [16]. The biggest challenge was to compensate intrinsic temperature coefficient of the ferrite field of -0.2% per °C. That was canceled down to the required 0.01%/°C by interspersing a thin NiFe "compensator alloy" strip between the ferrite bricks above and below the pole tips. The magnetic field drifted (logarithmically slow) by a minuscule 0.04% over many years of operation [17].

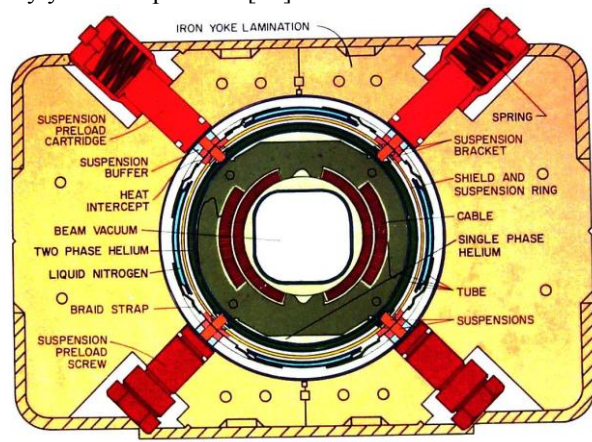

Figure 3: Cross-section of the Tevatron SC dipole magnet.

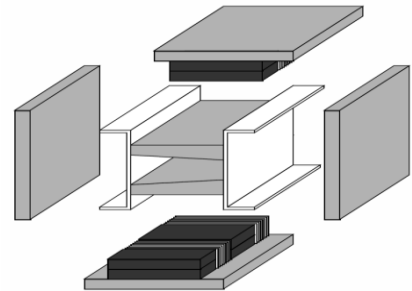

Figure 4: Recycler permanent magnet gradient dipole components shown in an exploded view. For every 4" wide brick there is a 0.5" interval of temperature compensator material composed of 10 strips.

*Production and Stochastic Cooling of Antiprotons :*
Stochastic cooling system technology at Fermilab has expanded considerably on the initial systems developed at CERN. A total of 25 independent cooling systems are utilized for increasing phase space density of 8 GeV antiprotons in three Fermilab antiproton synchrotrons: Accumulator, Debuncher, and Recycler. The development of the systems at Fermilab have been ongoing since the early days of commissioning in 1985, and greatly benefited from improvements of He-cooled pickup and kickers, preamplifiers, power amplifiers and recursive notch filters, better signal transmission and equalizers [18]. Together with progressed proton intensity on the target, better targetry, and electron cooling in Recycler, that led to stacking rates of antiprotons in excess of $28 \times 10^{10}$ per hour (world record – see Fig.5); stacks in excess of $300 \times 10^{10}$ have been accumulated in the Accumulator ring and $600 \times 10^{10}$ in the Recycler [19]. Very useful by-products of those developments were bunched beam stochastic cooling system, implemented at RHIC [20], and multi-GHz Schottky monitors successfully employed for multi-bunch non-invasive diagnostics of (simultaneously many) beam parameters Tevatron, Recycler and LHC [21] (see Fig.6).

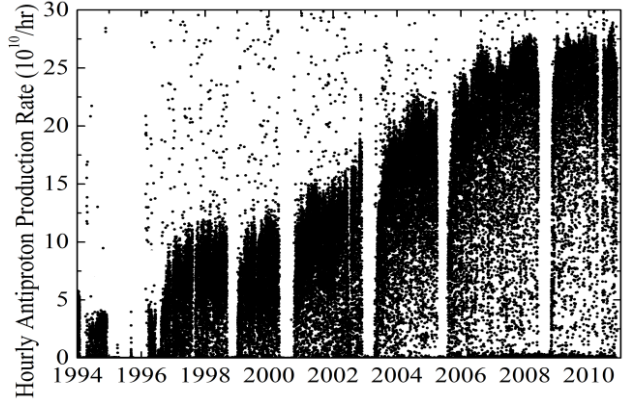

Figure 5: Average antiproton accumulation rate since 1994 and during all of Collider Run II (including production in the Antiproton Source and storage in RR).

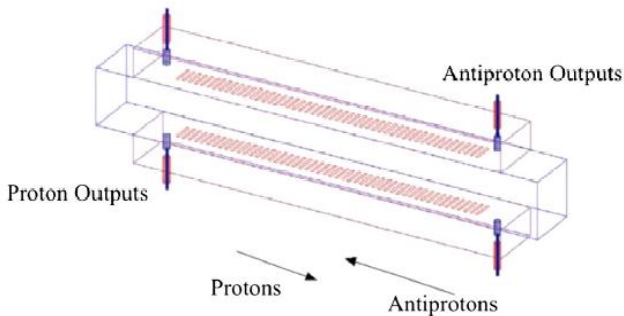

Figure 6: Slotted waveguide structure of 1.7 GHz Schottky monitors employed in the Tevatron.

*High Energy Electron Cooling of Antiprotons:*
One of the most critical elements in the evolution of Run II was the successful introduction of electron cooling [22] into the Recycler during the summer of 2005. Prior to the electron cooling luminosities had approached, but not exceeded, $1\times10^{32}$ cm$^{-2}$sec$^{-1}$, while the cooling opened the possibility for several times higher, record performance.

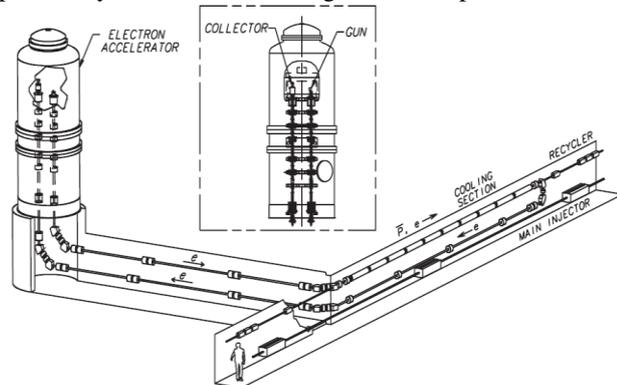

Figure 7: Schematic layout of the Recycler electron cooling system and accelerator cross section (inset) [22]

The project overcame not only the great challenge of operating 4.4 MeV Pelletron accelerator in the recirculation mode with upto 1A beams, but also resolved the hard issue of high quality magnetized beam transport through non-continuous magnetic focusing beamline [23].

*Slip-Stacking and Barrier-Bucket RF Manipulations:*
 Two innovative methods of longitudinal beam manipulation were been developed and implemented in operation and were crucial for the success of the Tevatron Run II: a) multi-batch slip stacking [24] that allowed to approximately double the 120 GeV proton bunch intensity for antiproton production; b) RF barrier-bucket system with rectangular 2kV RF voltage pulses [25] allowed for a whole new range of antiproton beam manipulation in the Recycler including operational "momentum mining" of antiprotons for the Tevatron shots [26] – see Figs.8 and 9.

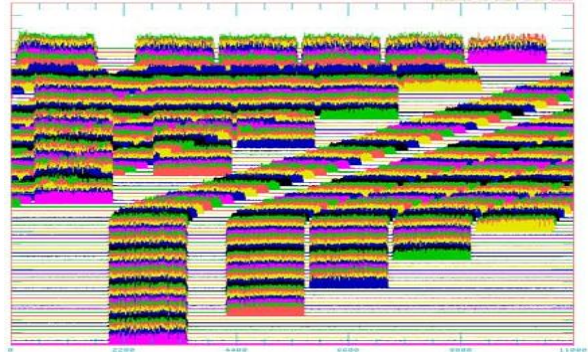

Figure 8: Mountain range plot showing 11 batch slip stacking process. Horizontal scale is 10 μsec.

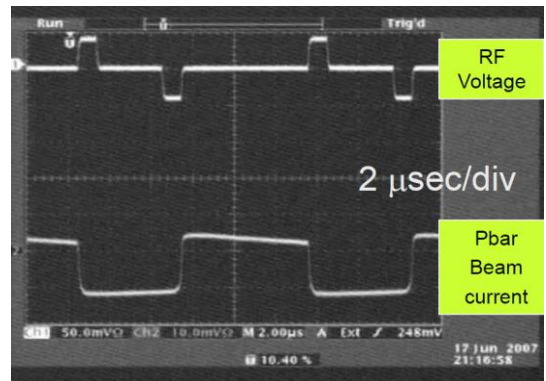

Figure 9: Recycler barrier bucket voltage (top) and beam profile (bottom) prior to "momentum mining".

*Electron Lenses for Beam-Beam Compensation:*
Electron lenses [27,28] are a novel accelerator technology used for compensation of the long-range beam-beam effects in the Tevatron [29,30], operational DC beam removal out of the Tevatron abort gaps [31], and, recently, for hollow electron beam collimation demonstration [32]. Two electron lenses were built and installed in A11 and F48 locations of the Tevatron ring. They use 1-3 A, 6-10 kV e-beam generated at the 10-15 mm diameter thermionic cathodes immersed in 0.3T longitudinal magnetic field and aligned onto (anti)proton beam orbit over about 2 m length inside 6T SC solenoid.

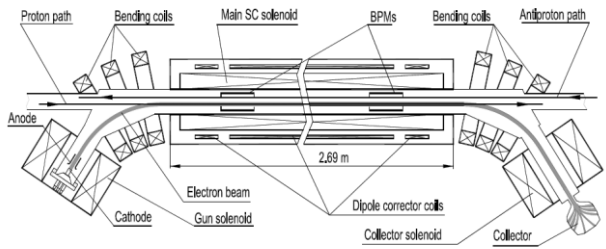

Figure 10: General layout of the Tevatron Electron Lens

We should emphasize, that the Tevatron Collider Runs not only delivered excellent performance (integrated luminosity), but also greatly advanced the whole accelerator field by studies of beam-beam effects [33], crystal collimation [34], electron cloud [35] and IBS [36], new theories of beam optics [37], IBS [38] and instabilities [39], sophisticated beam-beam and luminosity modeling [40] and more efficient beam instrumentation [41].

## LESSONS FROM TEVATRON

The Tevatron collider program will end on September 30, 2011. The machine has worked extremely well for 25 years. It has enabled CDF and D0 to discover the top quark and observe important features of the standard model for the first time. The Collider has greatly advanced accelerator technology and beam physics. Its success is a great tribute to the Fermilab staff.

We can draw several lessons from Tevatron's story:

a) one can see that exchange of ideas and methods and technology transfer helps our field: Fermilab scientists learned and borrowed a great deal of knowledge from ISR and SppS accelerators, and in turn, Tevatron's technology, techniques and experience have been successfully applied to HERA, RHIC and LHC;

b) operation of such complex systems as hadron colliders require us to be persistent and stubborn, pay close attention to details, do not count on "silver bullets" but instead be ready to go through incremental improvements.

c) It has taught us to be flexible, look for all possibilities to increase luminosity and not be afraid to change plans if experience shows the prospects diminishing due to the complexity of machines and often unpredictability of the performance limits. The expectations management is very important.

d) Operational difficulties not only generate strain, but also inspire and exalt creativity in the entire team of scientists and engineers, managers and technicians, support staff and collaborators. Hence many of us can say about the Tevatron years "it was the best time of my life".